\documentclass[pra,twocolumn,showpacs,amsmath,amssymb,eqsecnum,superscriptaddress]{revtex4-1}
\usepackage{graphicx,bm,color,mathptmx,hyperref,hhline} 
\usepackage{soul,bbm}
\newcommand{\Tr}{\mathop{\mathrm{Tr}} \nolimits}

\newcommand{\op}[1]{\hat{#1}}

\newcommand{\ket}[1]{\left\vert #1 \right\rangle}
\newcommand{\bra}[1]{\left\langle #1 \right\vert}

\begin{document}

\title{Unpolarized states and hidden polarization}

\author{P. de la Hoz} 
\affiliation{Departamento de \'Optica, Facultad de F\'{\i}sica, 
Universidad Complutense, 28040~Madrid, Spain}

\author{G.~Bj\"{o}rk} 
\affiliation{Department of Applied Physics,
  Royal Institute of Technology (KTH), AlbaNova,
  SE-106 91 Stockholm, Sweden}

\author{A.~B.~Klimov} 
\affiliation{Departamento de F\'{\i}sica,
  Universidad de Guadalajara, 44420~Guadalajara, Jalisco, Mexico}

\author{G.~Leuchs} 
\affiliation{Max-Planck-Institut f\"ur die Physik des Lichts, 
G\"{u}nther-Scharowsky-Stra{\ss}e 1, Bau 24, 91058 Erlangen, Germany} 
\affiliation{Department f\"{u}r Physik,
  Universit\"{a}t Erlangen-N\"{u}rnberg, Staudtstra{\ss}e 7, Bau 2,
  91058 Erlangen, Germany}

\author{L.~L.~S\'{a}nchez-Soto} 
\affiliation{Departamento de \'Optica, Facultad de F\'{\i}sica,
 Universidad Complutense, 28040~Madrid,  Spain} 
\affiliation{Max-Planck-Institut f\"ur die Physik des Lichts,
  G\"{u}nther-Scharowsky-Stra{\ss}e 1, Bau 24, 91058 Erlangen,
  Germany} 
\affiliation{Department f\"{u}r Physik, Universit\"{a}t
  Erlangen-N\"{u}rnberg, Staudtstra{\ss}e 7, Bau 2, 91058 Erlangen,
  Germany}

\begin{abstract}
  We capitalize on a multipolar expansion of the polarization density
  matrix, in which multipoles appear as successive moments of the
  Stokes variables. When all the multipoles up to a given order $K$
  vanish, we can properly say that the state is $K$th-order
  unpolarized, as it lacks of polarization information to that
  order.  First-order unpolarized states coincide with the
  corresponding classical ones, whereas unpolarized to any order tally
  with the quantum notion of fully invariant states. In between these
  two extreme cases, there is a rich variety of situations that are
  explored here. The existence of \textit{hidden} polarization emerges
  in a natural way in this context.
\end{abstract}

\pacs{42.25.Ja, 42.50.Dv, 42.50.Ar, 42.50.Lc}

\date{\today}

\maketitle

\section{Introduction}

Very often an involved physical concept can be better understood in
terms of its opposite.  Polarization is a pertinent example: perhaps
the most sensible way to look at it is to explore unpolarized states,
as one can make sense of them using exclusively invariance principles,
a tool of paramount importance in physics.
 
The constitution of unpolarized light was investigated from the very
beginning of modern optics.  Indeed, already
Stokes~\cite{Stokes:1852rt} and Verdet~\cite{Verdet:1869ve} offered a
lucid characterization of what they called ``natural'' light by using
the projections of the intensity onto the axes of a rotated Cartesian
coordinate system. Unpolarized states are those that remain invariant
under any rotation of that coordinate system and under any phase shift
between its rectangular components.

In classical optics, the field components of unpolarized light are
well modeled by zero-mean, uncorrelated, stationary Gaussian random
process~\cite{Kampen:2007qr}. The previous invariance conditions
thus determine the entire probabilistic structure of the projected
intensities~\cite{Barakat:1989fp}. However, as the standard 
theory is limited to first-order moments, unpolarized light is
presented as having zero-mean Stokes vector, which in geometrical
terms means that it is just the origin of the Poincar\'e
sphere~\cite{Born:1999yq}. We stress, though, that this is an
incomplete characterization, for it safely overlooks higher-order
moments~\cite{Paul:1994qq}.

At the quantum level, the invariance requirement fixes once and for
all the structure of the density matrix, as first pointed out in
Refs.~\cite{Prakash:1971fr,Agarwal:1971zr}: unpolarized states are
maximally mixed in each subspace with a given number of
photons~\cite{Lehner:1996rr,Soderholm:2001ay}.  To put it in another
way, it specifies the probability distribution and, as a result, all
the moments of the Stokes variables.

Nowadays, there is a widespread belief that a thorough appreciation of the
subtle effects arising in he quantum world requires a careful scrutiny
of higher-order polarization fluctuations.  We have been advocating
the use of a hierarchy of correlation functions that take into account
the successive moments of the Stokes variables~\cite{Muller:2012ys,
  Sanchez-Soto:2013cr,Hoz:2013om}. The most systematic way to
accomplish this is by expanding the density matrix in
multipoles~\cite{Blum:1981rb}. 

The idea of unpolarized states can be directly translated in this
scenario: when all the multipoles up to a given order (say $K$)
vanish, the state lacks of polarization information up to that order
and hence will be called $K$th-order unpolarized. The classical
picture matches the first-order theory, whereas the quantum condition
implies that all the multipoles are identically null.
Our goal here is to explore the \textit{terra incognita} between these
two extreme cases. In this respect, we mention that, as we shall see,
this is closely related with the so-called \textit{hidden}
polarization, introduced by Klychko~\cite{Klyshko:1992wd,Klyshko:1997yq}.

Our paper is organized as follows: In Sec.~\ref{Sec:Polstr} we
concisely sketch the fundamentals needed to grasp the polarization
hallmarks of quantum fields and introduce the multipoles.  In
Sec.~\ref{Sec:Korder} we revisit unpolarized states from the viewpoint
of these multipoles, defining $K$th unpolarized states. In
Sec.~\ref{Sec:Examples} we apply the formalism to some illuminating
examples and, finally, our conclusions are briefly summarized in
Sec.~\ref{Sec:Conc}.

\section{Polarization structure of quantum fields}
\label{Sec:Polstr}

A satisfactory description of the polarization structure of quantum
fields is of utmost significance for our purposes. This is precisely
the objective of this Section.

\subsection{The quantum polarization sector}

Let us consider a monochromatic field specified by two operators
$\op{a}_{H}$ and $\op{a}_{V}$, representing the complex amplitudes in
two linearly polarized orthogonal modes, we indicate as horizontal ($H$)
and vertical ($V$), respectively. The Stokes operators are~\cite{Luis:2000ys}
\begin{equation}
  \label{eq:Stokop}
  \begin{array}{c}
    \op{S}_{x} = \textstyle\frac{1}{2} 
    ( \op{a}^{\dagger}_{H}  \op{a}_{V} + 
    \op{a}^{\dagger}_{V} \op{a}_{H} ) \, ,  
    \qquad
    \op{S}_{y} =  \frac{i}{2} ( \op{a}_{H} \op{a}^{\dagger}_{V} - 
    \op{a}^{\dagger}_{H} \op{a}_{V} ) \, ,  \\
    \\
    \op{S}_{z}  = \frac{1}{2} ( \op{a}^{\dagger}_{H} \op{a}_{H} - 
    \op{a}^{\dagger}_{V} \op{a}_{V} ) \, ,
  \end{array}
\end{equation}
together with the total photon number
\begin{equation}
  \op{N} = \op{a}^{\dagger}_{H} \op{a}_{H} + 
  \op{a}^{\dagger}_{V} \op{a}_{V}  \, .
\end{equation}
The superscript $\dagger$ stands for the Hermitian adjoint.  In this
Schwinger representation~\cite{Schwinger:1965kx}, these operators
differ by a factor 1/2 from the common Stokes
parameters~\cite{Born:1999yq}, but in this way the components of the
Stokes vector $\op{\mathbf{S}} = (\op{S}_{x}, \op{S}_{y},
\op{S}_{z})$  satisfy the commutation relations of the
su(2) algebra:
\begin{equation}
  [ \op{S}_{x}, \op{S}_{y}] = i \op{S}_{z} \, ,
\end{equation}
and cyclic permutations (we use $\hbar =1$ throughout).

The noncommutability of these operators precludes the simultaneous
sharp measurement of the corresponding quantities. Among other
consequences, this implies that no field state (apart from the
two-mode vacuum) can have definite nonfluctuating values of all the
Stokes operators simultaneously. This is quantified by the uncertainty
relation
\begin{equation}
  \Delta^{2} \op{\mathbf{S}} =
  \Delta^{2} \op{S}_{x} + \Delta^{2} \op{S}_{y} + \Delta^{2} \op{S}_{z}
  \geq \textstyle{\frac{1}{2}} \langle \op{N} \rangle \, ,
  \label{eq:unrel}
\end{equation}
where $\Delta^{2} \op{S}_{j} = \langle \op{S}_j^2 \rangle - \langle
\op{S}_j \rangle^2$ are the variances. In this vein, one can say
that the electric vector of a monochromatic quantum field never
describes a definite ellipse.

Moreover, while the Stokes operators are all Hermitian, the
noncommutability makes mixed, nonsymmetric products (such as
$\hat{S}_{x} \hat{S}_{y}$) non-Hermitian, also precluding their direct
measurement.

In classical optics, the total intensity is a well-defined quantity
and the Poincar\'e sphere appears then as a smooth surface with radius
equal to that intensity.  In contradistinction, in quantum optics we
have
\begin{equation}
  \label{eq:nhmr}
  \op{\mathbf{S}}^{2} =
  \op{S}_{x}^{2}  + \op{S}_{y}^{2}  +  \op{S}_{z}^{2}  =
  S (S  + 1)  \op{\openone}   \, ,
\end{equation}
where $S =N/2$ plays the role of the spin ($N$ being the photon
number).  As fluctuations in $N$ are unavoidable (leaving aside
photon-number states), we are forced to talk of a three-dimensional
Poincar\'e space (with axis $S_{x}$, $S_{y}$ and $S_{z}$) that can be
envisioned as a set of nested spheres with radii proportional to the
different photon numbers that contribute significantly to the state.

We next make the important observation that
\begin{equation}
\label{eq:commNS}
  [ \op{N}, \op{\mathbf{S}} ] = 0 \, .
\end{equation} 
This expresses in the quantum language the fact that
polarization and intensity are separate concepts: the form of the
ellipse described by the electric field (polarization) does not depend
on its size (intensity). 

This fact brings about remarkable simplifications. First, it means
that each subspace with a fixed number of photons must be handled
separately. Equivalently, in the previous onion-like picture of the
Poincar\'e space, each shell has to be addressed independently.  This
can be emphasized if instead of the Fock states $\{ |n_H, n_V \rangle
\} $, which are an orthonormal basis of the Hilbert space of these
two-mode fields, we employ the relabeling
\begin{equation}
  | S, m \rangle \equiv | n_H = S + m, n_V = S - m \rangle \, ,
\end{equation}
which can be seen as the common eigenstates of $ \op{S}^{2}$ and
$\op{S}_{z}$.  For each fixed $S$, $m$ runs from $-S$ to $S$ and these
states span a $(2S+1)$-dimensional invariant subspace, wherein the
operators $\op{\mathbf{S}}$ act in the standard form
\begin{eqnarray}
  \op{S}_{\pm} \, | S, m \rangle & = &
  \sqrt{ S (S+1 ) -  m (m \pm 1) } \, |S, m \pm   1 \rangle \, ,
  \nonumber \\ 
  & & \\
  \op{S}_{z} \, | S, m \rangle & = &
  m | S, m \rangle \, . \nonumber
\end{eqnarray}

Second, from \eqref{eq:commNS} it follows that any function of the
Stokes operators $f (\op{\mathbf{S}} )$ commutes with
$\op{N}$. Therefore, the matrix elements of the density matrix
$\op{\varrho}$ (which describes the state) connecting subspaces with
different values of $S$ do not contribute to $\langle f
(\op{\mathbf{S}} ) \rangle$. Put differently, the only accessible
polarization information from any state $\op{\varrho}$ is its
block-diagonal form
\begin{equation}
  \label{eq:rhopol}
  \op{\varrho}_{\mathrm{pol}} = \bigoplus_{S}  P_{S} \;
  \op{\varrho}^{(S)}  \, ,
\end{equation}
where $P_{S}$ is the photon-number distribution ($S = 0, 1/2, 1,
\ldots$) and $P_{S} \, \op{\varrho}^{(S)}$ is the reduced density
matrix in the subspace with spin $S$. This
$\op{\varrho}_{\mathrm{pol}}$ has been termed the polarization
sector~\cite{Raymer:2000zt} or the polarization density
matrix~\cite{Karassiov:2004xw}. What matters for our purposes is that
any $\op{\varrho}$ and its associated $\op{\varrho}_{\mathrm{pol}} $
cannot be distinguished in polarization measurements and, accordingly,
we shall be using the block-diagonal form \eqref{eq:rhopol} and drop
henceforth the subscript pol.

\subsection{Polarization multipoles}

Instead of using directly the states $\{ | S, m \rangle \}$, it
is more convenient  to expand each component
$\op{\varrho}^{(S)}$ in \eqref{eq:rhopol} as
\begin{equation}
  \label{rho1}
  \op{\varrho}^{(S)} =  \sum_{K= 0}^{2S} \sum_{q=-K}^{K}  
  \varrho_{Kq}^{(S)} \,   \op{T}_{Kq}^{(S)} \, .
\end{equation}
The irreducible tensor operators $T_{Kq}^{(S)}$
are~\cite{Varshalovich:1988ct}
\begin{equation}
  \label{Tensor} 
  \op{T}_{Kq}^{(S)} = \sqrt{\frac{2 K +1}{2 S +1}} 
  \sum_{m,  m^{\prime}= -S}^{S} C_{Sm, Kq}^{Sm^{\prime}} \, 
  |  S , m^\prime \rangle \langle S, m | \, ,
\end{equation}
with $ C_{Sm, Kq}^{Sm^{\prime}}$ being the Clebsch-Gordan coefficients
that couple a spin $S$ and a spin $K$ \mbox{($0 \le K \le 2S$)} to a
total spin $S$.  These tensors are an orthonormal basis
\begin{equation}
  \Tr  [ \op{T}_{K q}^{(S)} \, 
  \op{T}_{K^{\prime} q^{\prime}}^{(S^{\prime}) \, \dagger}   ] =
  \delta_{S S^{\prime}}  \delta_{K  K^{\prime}} \delta_{q q^{\prime}} \, ,
\end{equation}
and they have the right transformation properties: under a rotation
parametrized by the Euler angles $(\alpha, \beta, \gamma)$, we have
\begin{equation}
  \op{R} (\alpha , \beta, \gamma) \, \op{T}_{Kq}^{(S)}  \,
  \op{R}^{\dagger} (\alpha , \beta, \gamma) =
  \sum_{q^{\prime}}  D_{q^{\prime} q}^{S} (\alpha, \beta, \gamma ) \,
  \op{T}_{Kq^{\prime}}^{(S)}  \, ,
  \label{eq:sym3}
\end{equation}
where the $D_{q^{\prime} q}^{S} (\alpha, \beta, \gamma) $ stands for
the matrix elements of the rotation operator $\op{R} (\alpha, \beta,
\gamma)$ in the basis $ |S, m \rangle$~\cite{Varshalovich:1988ct}.

Although at first sight they might look a bit intimidating,
they are nothing but the multipoles used in atomic
physics~\cite{Blum:1981rb}.  After some calculations, one can recast
Eq.~\eqref{Tensor} as
\begin{equation}
  \label{eq:multi}
  \begin{array}{l}
    \op{T}_{00}^{( S )}  =  
    \frac{1}{\sqrt{2 S + 1}} \op{\openone} \, , \\
    \\
    \op{T}_{10}^{( S )}  =   \textstyle{\sqrt{\frac{3}{( 2 S + 1 ) (S+1) S}}} \, 
    \op{S}_{z} \, , \quad  
    \op{T}_{1\mp 1}^{( S )}  =   \sqrt{\frac{3}{( 2 S + 1 ) (S+1) S}}  \, 
    \op{S}_{\pm} \, , \\
    \\
    \op{T}_{20}^{{(S}} = \textstyle{\sqrt{\frac{C}{6}}}  (3\op{S}_{z}^{2}-
    \op{S}^{2}) \, , \ \
    \op{T}_{2\mp 1}^{( S )}  =   \textstyle{\sqrt{\frac{C}{2}}} \, 
    \{ \op{S}_{z},  \op{S}_{\pm} \}  \, , \ \
    \op{T}_{1\mp 2}^{( S )}  =   \textstyle{\sqrt{\frac{C}{2}}}   \, 
    \op{S}_{\pm}^{2} \, ,
  \end{array}
\end{equation}
where $C=30/[(2S + 3)(2 S + 1) (2S-1) (S+1)]$ and $\{ , \}$ is the
anticommutator.  In consequence, we conclude that $\op{T}^{(S)}_{Kq}$
can be related to the $K$th power of the Stokes
operators~(\ref{eq:Stokop}).

The corresponding expansion coefficients
\begin{equation}
  \varrho_{Kq}^{(S)} =  \Tr [ \op{\varrho}^{(S)} \,
  \op{T}_{Kq}^{(S) \, \dagger} ]
\end{equation}
are known as state multipoles. The hermiticity imposes the symmetry
condition
\begin{equation}
  \label{eq:1}
  \varrho_{K-q}^{(S)} = ( -1)^{q} \, \varrho_{Kq}^{(S)} \, ,
\end{equation}
and the positive semidefiniteness of $\op{\varrho}^{(S)}$ forces the bound
\begin{equation}
  \mathcal{W}_K^{(S)} \equiv 
  \sum_{q=- K}^{K} |\varrho_{Kq}^{(S)}|^{2} \leq  C_{K}^{(S)} \, ,
  \label{eq:cons}
\end{equation}
for every $K>1$ and $ C_{K}^{(S)}$ a positive constant.  The quantity
$\mathcal{W}_K^{(S)} $ is just the strength of the $K$th multipole in
the state  $\op{\varrho}^{(S)}$.

Finally, we turn to the important class of axially symmetric
states~\cite{Blum:1981rb}. They are invariant under rotations about an
axis that we take as the $z$ axis. Since $ D_{qq^{\prime}}^{S}
(0,0,\gamma) = \exp(-i q \gamma) \delta_{qq^{\prime}}$, this implies
\begin{equation}
  \op{\varrho}_{\mathrm{ax  sym}}^{(S)} = 
  \sum_{K=0}^{2S} \varrho_{K0}^{(S)}  \, \op{T}_{K0}^{(S)}   \, .
  \label{eq:axsys}
\end{equation}
Thus, they are characterized exclusively by the multipole components
$\varrho_{K0}$.  Any density operator that can be obtained from $
\op{\varrho}_{\mathrm{axsym}} $ via an SU(2) transformation, 
represents as well an axial symmetric state, as a rotation only change the
direction of the symmetry axis of the state.

Some axially symmetric systems are also invariant under the reversal
of the symmetry axis (i.e., $z \rightarrow -z$). As this corresponds
to a rotation around the $y$ axis by an angle $\pi$ and $
D_{qq^{\prime}}^{S} (0, \pi , 0) = (-1)^{K+q} \delta_{q \, -
  q^{\prime}}$, we get from \eqref{eq:sym3}
\begin{equation}
  \label{eq:invz}
  \varrho_{K0}^{(S)} = (-1)^{K} \varrho_{K0}^{(S)} \, ,
\end{equation}
so only multipoles of even rank $K$ contribute.

\section{$K$th-order unpolarized states}
\label{Sec:Korder}
 
From now on, we restrict ourselves to a single component
$\op{\varrho}^{(S)}$ (i.e., a fixed number of photons). This is by no
means a restriction, as the discussion can be extended in a natural
way to the complete polarization density matrix in
\eqref{eq:rhopol}.

The full polarization information is encoded in the complete multipole
distribution $\{\mathcal{W}_{K}^{(S)} \}$.  However, for most of the
states, only a limited number of multipoles play a substantive role
and the rest of them have a small contribution.  As a consequence,
gaining a good feeling of the corresponding behavior may be tricky.

A way to bypass this disadvantage is to look at the cumulative
distribution
\begin{equation}
  \label{eq:cum}
  \mathcal{A}^{(S)}_{K} = \sum_{\ell = 1}^{K} \mathcal{W}^{(S)}_{\ell}   \, .
\end{equation}
Please, note carefully that the monopolar term has been
excluded, as it is trivially isotropic for all the states. The
quantity $\mathcal{A}^{(S)}_{K}$ conveys whole information \textit{up}
to order $K$.  We know from probability that it has
remarkable properties~\cite{Jaynes:2003uq}. 

To get extra insights into this subject, let us focus, for the time
being, on the key example of SU(2) coherent states (also known
as spin or atomic coherent states), which can be written down as the
superposition~\cite{Arecchi:1972zr,Perelomov:1986ly}
\begin{equation}
  \label{eq:psiN}
  | \theta, \phi \rangle = \sum_{m=-S}^{S}  
  C_{Sm} (\theta, \phi)  \, | S, m \rangle \, ,
\end{equation}
with coefficients
\begin{equation}
  C_{Sm} (\theta, \phi)  =
  \left (
    \begin{array}{c}
      2S \\
      S+m
    \end{array}
  \right )^{1/2}
  \left ( \sin \frac{\theta}{2} \right)^{S+m}
  \left ( \cos \frac{\theta}{2} \right)^{S-m}
  e^{-i (S+m) \phi} \, ,
\end{equation}
and $\theta$ and $\phi$ are the polar and azimuthal angles on the
sphere, respectively. If $\mathbf{n}$ is the unit vector in the
direction~$(\theta, \phi)$, the operator $ \op{S}_{\mathbf{n}} =
\mathbf{n} \cdot \op{\mathbf{S}}$ is the observable measured in
polarization experiments~\cite{Marquardt:2007bh}: coherent states can
be alternatively interpreted as eigenstates of $\op{S}_{\mathbf{n}}$
\begin{equation}
  \label{eq:CSe}
  \op{S}_{\mathbf{n}}  |\theta , \phi \rangle = 
 S |\theta , \phi  \rangle \, ,
\end{equation}
whence one can confirm that hey saturate the uncertainty relation
(\ref{eq:unrel}) and so they are the minimum uncertainty states in
polarization optics.
 
For these states, one can immediately find
\begin{equation}
  \label{eq:Aksu2}
  \mathcal{A}^{(S)}_{K,\mathrm{SU(2)}} = \frac{2S}{2S +1} -
  \frac{[\Gamma (2S + 1)]^{2}}{\Gamma (2S-K) \Gamma (2S + K +2)} \, .
\end{equation}
As conjectured in Ref.~\cite{Hoz:2013om}, this is the maximal value
attainable for any $K$ in each subspace $S$. This nicely corroborates the
amazing properties of  SU(2) coherent states:  they are maximally
polarized states to any order, as one would expect from a pure state
that corresponds as nearly as possible to a classical spin vector
pointing in a given direction. 

This maximal property suggests at once a hierarchy of degrees of
polarization
\begin{equation}
  \label{eq:PK}
  \mathbb{P}_{K}^{(S)} =
  \sqrt{\frac{\mathcal{A}^{(S)}_{K}}{\mathcal{A}^{(S)}_{K,\mathrm{SU(2)}}}} \, ,
\end{equation}
which sort the relevant polarization information \textit{up} to order
$K$~($K= 1 , \ldots, S$).  The experimental
reconstruction reported in Ref.~\cite{Hoz:2013om} reveals that by
performing a Stokes measurement in $2K+1$ independent directions, one
can actually determine $\mathcal{A}^{(S)}_{K}$ and hence all the
degrees  $\mathbb{P}_{K}^{(S)}$.

We will say that a state is $K$th-order unpolarized when
$\mathbb{P}_K^{(S)}=0$ (which obviously implies
$\mathcal{A}_K^{(S)}=0$; i.e., all the multipoles up to order $K$
vanish) and we will denote these states as
$\op{\varrho}_{\mathrm{unpol},K}^{(S)}$. Note, though,  that
$K$th-order unpolarized states do carry polarization information when
one looks at higher-order moments. This is referred to as
\textit{hidden} polarization, according to the terminology coined by
Klyshko~\cite{Klyshko:1992wd,Klyshko:1997yq}, albeit it would be
better to say that such states display higher-order
polarization~\cite{Gupta:2011lq}.

In more physical terms, the condition of $K$th-order unpolarization
amounts to imposing that the moments $\langle
\op{\mathbf{S}}_{\mathbf{n}}^{\ell} \rangle$ are independent of the
direction $\mathbf{n}$ for $\ell = 1, \ldots, K$ (i.e., they are
isotropic). Therefore, all the moments up to order $K$ do not show any
angular structure, whereas higher-order ones do. Notice, in passing,
that this is the logic beyond the recent proposal of
anti-coherent states~\cite{Zimba:2006fk}: such states ``point
nowhere'' (to a given order), and this is certainly one way to serve as the
opposite of a state that points, as much as possible, somewhere.  From
this perspective, these unpolarized states are exhibits the most
nonclassical features~\cite{Giraud:2010db}.

For the particular case of the dipole ($K=1$), Eq.~(\ref{eq:PK})
reduces to
\begin{equation}
  \mathbb{P}_{1}^{(S)}= 
  \frac{ \sqrt{\langle \op{S}_{x} \rangle^2+ \langle \op{S}_{y} \rangle^2 +
      \langle \op{S}_{z} \rangle^2}}{S} \, ,
\end{equation}
which coincides with the standard definition~\cite{Bjork:2010rt}.
First-order unpolarized states verify $\mathbb{P}_{1}^{(S)} = 0$, so
$\langle \op{\mathbf{S}} \rangle = 0$. This goes to the classical
notion of random states, as it involves exclusively first-order
moments.   

When the state is unpolarized to any order, only the monopole
contributes:
\begin{equation}
  \label{eq:unp}
  \op{\varrho}_{\mathrm{unpol}}^{(S)} =\frac{1}{2S+1} \,
  \openone_{2S+1} \, ,
\end{equation}
so it is a random state within each invariant
subspace. This is the quantum
definition, which demands that the whole probability
distribution to be SU(2)
invariant~\cite{Prakash:1971fr,Agarwal:1971zr}; that is, 
\begin{equation}
  \label{eq:2}
  [\op{\varrho}, \op{\mathbf{S}} ] = 0 \, ,
\end{equation}
wherefrom Eq.~\eqref{eq:unp} follows~\cite{Soderholm:2001ay}.  The
vacuum state is the only pure state that is unpolarized according to
this definition, and unpolarized mixed states are maximally
mixed in each subspace $S$. Any two-mode thermal state is hence
unpolarized.

  \section{A menagerie of unpolarized states}
  \label{Sec:Examples}

\subsection{Single-photon unpolarized states}

Single-photon states ($S=1/2$) are fairly special: they can only be
first-order unpolarized.  The multipole expansion of a general
single-photon state reads
\begin{equation}
  \op{\varrho}^{(1/2)} = \varrho_{00}^{(1/2)} \, \op{T}_{00}^{(1/2)}   + 
  \sum_{q}  \varrho_{1q}^{(1/2)} \op{T}_{1q}^{(1/2)} \, .
\end{equation} 
Since the state has only dipolar component, quantum and classical
descriptions coincide.  Positivity constraints the posible values of
the dipole to the range $ 0 \leq \mathcal{A}_{1}^{(1/2)} \leq 1/2$.
The condition $\mathcal{A}_{1}^{1/2} =0$ fixes at once unpolarized
states; viz,
\begin{equation}
  \varrho_{\mathrm{unpol},1}^{(1/2)} = 
  \frac{1}{2} \; 
  \left (
    \begin{array}{cc}
      1 & 0 \\
      0 & 1
    \end{array}
  \right ) \, .
\end{equation}
These states are both classically and quantum unpolarized, but, like
all quantum objects, can only be considered as elements of an
ensemble~\cite{Peres:2002oz}.

\subsection{Two-photon unpolarized states}

For two-photon states, there are first-order (or classical) and
second-order (or quantum) unpolarized states. The general condition
for first-order unpolarization is
\begin{equation}
  \op{\varrho}_{\mathrm{unpol},1}^{(1) } = 
  \varrho_{00}^{(1)} \, \op{T}_{00}^{(1)} 
  +\sum_{q} \varrho_{2 q}^{(1)} \, \op{T}_{2 q}^{(1)} \, ,
  \label{eq:un2ph}
\end{equation}
with the extra constraint of positivity.

  \begin{figure}[b]
    \includegraphics[width=.85\columnwidth]{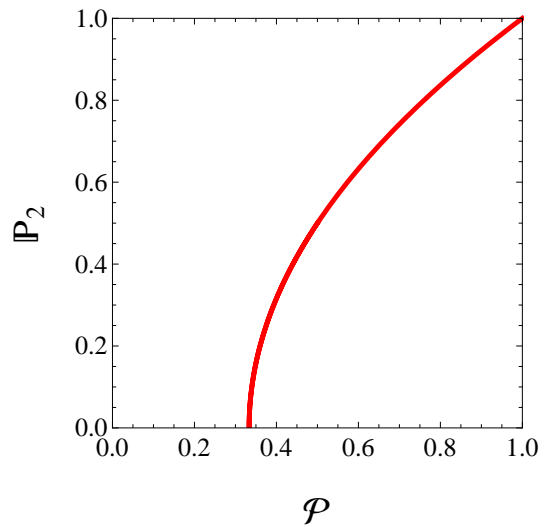}
    \caption{Second-order  degree of polarization as a function of the
      purity, for the first-order unpolarized states \eqref{eq:diag2}.}
    \label{hist}
  \end{figure}

Special attention deserves the case of axially symmetric states as
they can always be diagonalized via SU(2) rotations; viz,
$\op{\rho}_{\mathrm{as sym}}^{(1)} = \op{R} (\alpha,\beta,\gamma) \,
\op{\varrho}_{d}^{(1)} \, \op{R}^{\dagger } (\alpha,\beta,\gamma)$, with
\begin{eqnarray}
  \op{\varrho}_{d}^{(1)} & = & 
  \left(
    \begin{array}{ccc}
      \lambda_{1} & 0 & 0 \\
      0 & \lambda_{2} & 0 \\
      0 & 0 & \lambda_{3}
    \end{array}
  \right)  
\nonumber \\
  & =& 
  \frac{1}{\sqrt{3}}  \op{T}_{00}^{(1)} +
  \frac{\lambda_{1}-\lambda_{3}}{\sqrt{2}} \op{T}_{10}^{(1)} + 
  \frac{1-3\lambda_{2}}{\sqrt{6}} \op{T}_{20}^{(1)} \, .
  \label{eq:diag}
\end{eqnarray}
The state is first-order unpolarized when $\lambda_{1}
=\lambda_{3}$. Since $\Tr (\op{\varrho}_{d}) = 1$, we can write
\begin{equation}
  \op{\varrho}^{(1)}_{\mathrm{unpol},1} = 
  \left ( 
    \begin{array}{ccc}
      \lambda & 0 & 0\\
      0 & 1-2\lambda & 0\\
      0 & 0 & \lambda
    \end{array}
  \right ) \, ,
  \label{eq:diag2}
\end{equation}
and positivity enforces $ 0 \leq \lambda \leq 1/2$, i.e., $0 \leq
\mathcal{A}_2^{(1)} \leq 2/3$. Both the purity $\mathcal{P}^{(1)} =
\Tr \{ [\op{\varrho}_{d}^{(1)}]^{2} \} $ and the second-order degree $
\mathbb{P}_{2}^{(1)} $ depend on $\lambda$
  \begin{equation}
   \mathcal{P}^{(1)} = 6 \lambda^{2} - 4 \lambda + 1 \, , 
\qquad
 \mathbb{P}_{2}^{(1)} =\sqrt{(3 \lambda -1)^2} \, ,
  \end{equation}
while $\mathbb{P}_{1}^{(1)} =0$ as anticipated. This can be recast as
\begin{equation}
  \mathbb{P}_{2}^{(1)} =\sqrt{[3 \mathcal{P}^{(1)} -1]/2} \, .
\end{equation}

In Fig.~\ref{hist} we have plotted $\mathbb{P}_{2}^{(1)}$ as a
function of the purity.  The maximum degree $\mathbb{P}_{2}^{(1)}$ is
attained for the pure states
\begin{equation}
  \label{eq:pSon}
  |\Psi_{\mathrm{unpol,1}}^{(1)} \rangle =  \frac{1}{\sqrt{2}} 
  \sin \beta    [ e^{i\alpha} |1, 1 \rangle - e^{-i\alpha}
  |1,-1\rangle ] + \cos \beta  |1,0\rangle \, ,
\end{equation}
and they are the transformed of the state $|1,0 \rangle$ under SU(2)
rotations $\op{R}(\alpha, \beta, \gamma)$.  Incidentally, these states
have served as the thread to experimentally verify the existence of
\textit{hidden} polarization~\cite{Usachev:2001ve,Sehat:2005wd}. They
coincide with the anticoherent states introduced in
Ref.~\cite{Zimba:2006fk} and worked out using the Majorana
representation~\cite{Crann:2010qd,Bannai:2011pi}. Unfortunately, their
nice geometric properties cannot be extended to mixed
states~\eqref{eq:diag2}.

\subsection{Three-photon unpolarized states}

For three-photon states we have first- (classical), second- and
third-order (quantum) unpolarized states.  

\begin{figure}
  \includegraphics[width=.37\columnwidth]{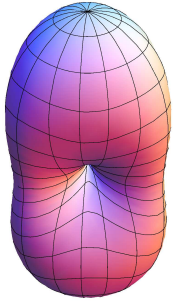}
  \caption{$Q$-function for three-photon first-order pure unpolarized
    states~\eqref{eq:3p}.}
  \label{fig:fig3}
\end{figure}

\begin{figure}
  \begin{center}
    \includegraphics[width=0.49\linewidth]{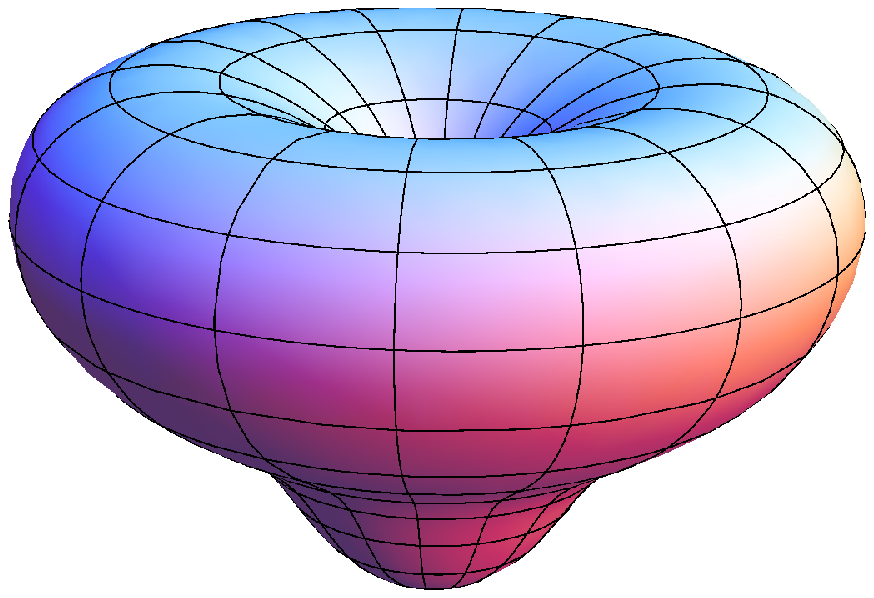}
    \hfill
    \includegraphics[width=0.49\linewidth]{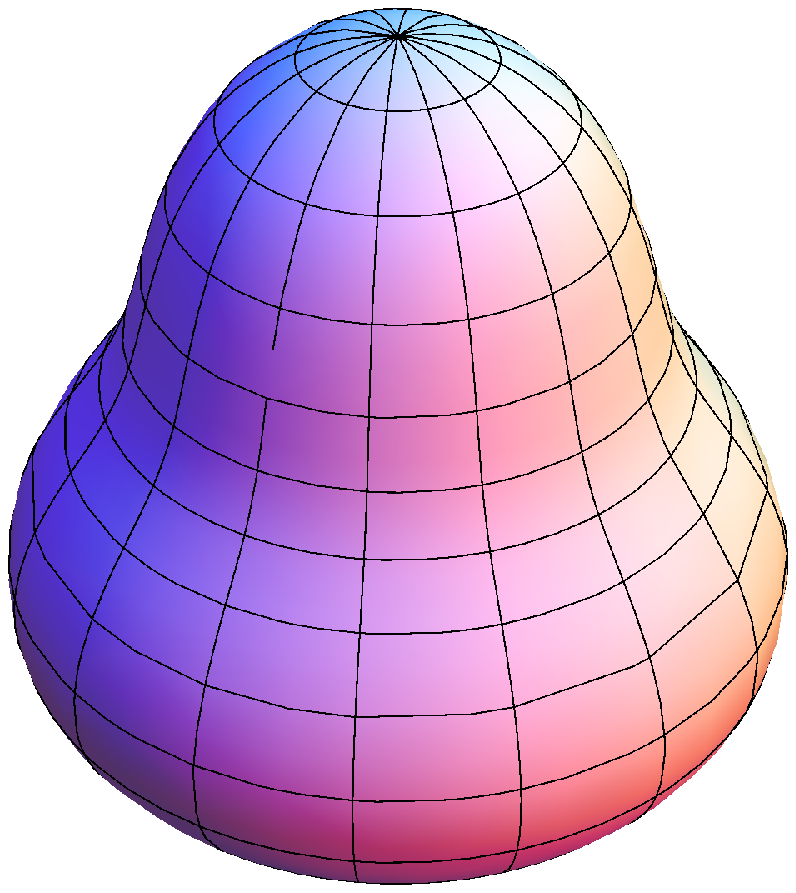}
  \end{center}
  \caption{$Q$-function for axially symmetric three-photon
    second-order unpolarized states with maximal purity. In the left,
    we represent the state  $3/4 \ket{3/2,1/2}
   \bra{3/2,1/2}+1/4\ket{3/2,-3/2}\bra{3/2,-3/2}$,
    while in the right the state
    $1/3\ket{3/2,3/2}\bra{3/2,3/2}+1/2 
   \ket{3/2,-1/2}\bra{3/2,-1/2}+1/6\ket{3/2,-3/2}\bra{3/2,-3/2}$
    is plotted.}
  \label{fig:fig4}
\end{figure}

The diagonalizable states can be brought to the form
\begin{eqnarray}
  \op{\varrho}_{d}^{(3/2)} & =& \left (
    \begin{array}{cccc}
      \lambda_{1} & 0 & 0 & 0 \\
      0 & \lambda_{2} & 0 & 0 \\
      0 & 0 & \lambda_{3} & 0 \\
      0 & 0 & 0 & \lambda_{4} \\
    \end{array}
  \right )  =     \frac{1}{2} \op{T}_{00}^{(3/2)} 
\nonumber\\
& - &
\left(\frac{2 \lambda _2+4 \lambda _3+6 \lambda _4-3}{2 \sqrt{5}}\right)
  \op{T}_{10}^{(3/2)} 
   +  \textstyle{\left(\frac{1}{2} -\lambda _2-\lambda _3\right)} 
 \op{T}_{20}^{(3/2)} \nonumber  \\
& + &
  \left(\frac{-4 \lambda _2+2 \lambda _3-2 \lambda _4+1}{2 \sqrt{5}}\right)
  \op{T}_{30}^{(3/2)} \, . 
\end{eqnarray}
The condition for first-order unpolarization is
\begin{equation}
  2 \lambda_2+4 \lambda_3+6 \lambda_4-3 = 0 \, ,
\end{equation}
and the dipole is absent.  Now there are not axially-symmetric
first-order unpolarized states, because when $S$ is a half-integer, no
state of the basis lacks first-order polarization.

The diagonal operator of a three-photon first-order unpolarized state
reads
\begin{equation}
  \op{\varrho}_{\mathrm{unpol},1}^{(3/2)} =\left(
    \begin{array}{cccc}
      \lambda_3 +2 \lambda_4 -1/2 & 0 & 0 & 0 \\
      0 & -2 \lambda_3-3 \lambda_4+3/2 & 0 & 0 \\
      0 & 0 & \lambda_3 & 0 \\
      0 & 0 & 0 & \lambda_4 \\
    \end{array}
  \right) \, ,
  \label{eq:3pho}
\end{equation} 
and positivity translates into $ 0 \leq \lambda_3+2 \lambda_4-1/2 \leq
1$ and $ 0 \leq -2 \lambda_3 -3 \lambda_4 +3/2 \leq 1$,  $0 \leq
\lambda_3, \lambda_4 \leq 1$. The purity $\mathcal{P}$ is
\begin{equation}
  \mathcal{P}=  \frac{1}{4}+
  \frac{5}{4} \left(2 \lambda_3+2
    \lambda_4-1\right){}^2+
  \left(\lambda_3+3 \lambda_4-1\right){}^2 \, .
\end{equation}
with the bounds $1/4 \leq \mathcal{P} \leq 5/8$.

We remark that by using the Majorana representation mentioned above,
one can check that the SU(2) transformed of
\begin{equation}
  |\Psi_{\mathrm{unpol},1}^{(3/2) }\rangle = 
  \frac{1}{\sqrt{2}} |3/2, -3/2 \rangle +
  \frac{1}{\sqrt{2}} |3/2,3/2 \rangle \, ,
  \label{eq:3p}
\end{equation}
are first-order unpolarized, although they do not fall in the class of
states defined by~\eqref{eq:3pho}. To better appreciate these states,
one can work out the SU(2) $Q$ function, which is defined
as~\cite{Stratonovich:1956qc,Berezin:1975mw}
\begin{equation}
  \label{eq:QSU2j}
  Q^{(S)} (\theta, \phi) = \langle  \theta, \phi |
  \op{\varrho}^{(S)}  | \theta, \phi  \rangle \, ,
\end{equation}
where $ | \theta, \phi \rangle $ are SU(2) coherent states. In Fig.~2
we plot this $Q$ function for the state \eqref{eq:3p}.

To get a second-order unpolarized state, we need the additional
condition: $\lambda _3+3 \lambda _4-1=0$ and, the diagonal form for
these states is
\begin{equation}
  \op{\varrho}_{\mathrm{unpol,2}}^{(3/2)} = 
  \left(
    \begin{array}{cccc}
      \frac{1}{2}-\lambda _4 & 0 & 0 & 0 \\
      0 & 3 \lambda _4-\frac{1}{2} & 0 & 0 \\
      0 & 0 & 1-3 \lambda _4 & 0 \\
      0 & 0 & 0 & \lambda _4 \\
    \end{array}
  \right) \, . 
  \label{eq:diag32nd}
\end{equation}
The maximal purity of a second-order unpolarized axially symmetric
three-photon state is $\mathcal{P}=7/18$. In Fig.~\ref{fig:fig4}, we
represent the $Q$ function for second-order unpolarized states
maximizing the purity.

  \section{Concluding remarks}
  \label{Sec:Conc}

  Multipolar expansions are a powerful machinery.  We have applied
  such an expansion to the polarization density matrix, showing how
  the multipoles quantify higher-order fluctuations in the Stokes
  variables. In this way, we have provided a systematic
  characterization of unpolarized states as those states whose
  multipoles up to a given order vanish.

  The formalism can be extended to systems in which SU(2) symmetry
  plays a crucial role (such as Bose-Einstein condensates, spin
  chains, etc) and to other unitary symmetries, such as SU(3) (which
  is pivotal to understanding the polarization of the near field).

\begin{acknowledgments}
  Financial support from the Swedish Foundation for International
  Cooperation in Research and Higher Education (STINT), the Swedish
  Research Council (VR) through its Linn{\ae}us Center of Excellence
  ADOPT and contract No.~621-2011-4575, the Mexican CONACyT (Grant
  No.~106525), the EU FP7 (Grant Q-ESSENCE), and the Spanish MINECO
  (Grant FIS2011-26786) is gratefully acknowledged. 
\end{acknowledgments}


%

\end{document}